\pdfoutput=1%
\documentclass[a4paper,american,floatfix,pdftex,superscriptaddress,twoside,%
aip,jcp,%
citeautoscript,
longbibliography,%
reprint,twocolumn,final
]{revtex4-1}%
\usepackage{amsfonts,amsmath,amssymb}
\usepackage[T1]{fontenc}
\usepackage{graphicx}%
\usepackage[utf8]{inputenc}
\usepackage{microtype}
\usepackage{titlesec}
\usepackage[obeyFinal,textsize=scriptsize]{todonotes}
\usepackage{xspace}
\usepackage{hyperref, hypernat}
\usepackage[todos]{CMImacros}

\graphicspath{{./fig/}} 
\titleformat{\paragraph}[runin]{\bfseries\vspace{1em}}{}{1em}{\hspace{-\parindent}}

\makeatletter%
\renewcommand*{\@fnsymbol}[1]{\ensuremath{\ifcase#1\or *\or \|\or **\or \mathparagraph\or
      \mathsection\or \dagger\or \ddagger\or \dagger\dagger \or \ddagger\ddagger \else\@ctrerr\fi}}
\makeatother

\newcommand{\cfeldesy}{\affiliation{Center for Free-Electron Laser Science, Deutsches
      Elektronen-Synchrotron DESY, Notkestrasse 85, 22607 Hamburg, Germany}}%
\newcommand{\uhhcui}{\affiliation{The Hamburg Center for Ultrafast Imaging, Universität Hamburg,
      Luruper Chaussee 149, 22761 Hamburg, Germany}}%
\newcommand{\uhhphys}{\affiliation{Department of Physics, Universität Hamburg, Luruper Chaussee 149,
      22761 Hamburg, Germany}}%
\newcommand{\granada}{\affiliation{Instituto Carlos I de F\'{\i}sica Te\'orica y Computacional and
      Departamento de F\'{\i}sica At\'omica, Molecular y Nuclear, Universidad de Granada, 18071
      Granada, Spain}}%
\newcommand{\ayemail}{\email[]{andrey.yachmenev@cfel.de}}%
\newcommand{\rgfemail}{\email[]{rogonzal@ugr.es}}%
\newcommand{\jkemail}{\email[Email: ]{jochen.kuepper@cfel.de}}%
\newcommand{\cmiweb}{\homepage[\\website: ]{https://www.controlled-molecule-imaging.org}}%

\begin{document}
\title{The Effect of Nuclear-Quadrupole Coupling in the Laser-Induced Alignment of Molecules}
\author{Linda V. Thesing}\cfeldesy\uhhcui\uhhphys%
\author{Andrey Yachmenev}\ayemail\cfeldesy\uhhcui%
\author{Rosario Gonz{\'a}lez-F{\'e}rez}\rgfemail\granada%
\author{Jochen Küpper}\jkemail\cmiweb\cfeldesy\uhhcui\uhhphys%
\date{\today}%
\begin{abstract}\noindent%
   \textbf{Abstract:} We present a theoretical study of the time-dependent laser alignment of
   molecules taking into account the hyperfine coupling due to nuclear-quadrupole interactions. The
   coupling of nuclear spins to the overall angular momentum of molecules significantly influences
   their rotational dynamics. Here, we systematically analyze the impact of the nuclear-quadrupole
   coupling on the rotational dynamics of the linear \Itwo and the asymmetric-top diiodobenzene
   molecule induced by external laser fields. We explore different regimes of pulse shapes and
   laser-pulse intensities and detail under which conditions the quadrupole coupling cannot be
   neglected in the description of the laser alignment of molecules.
\end{abstract}
\maketitle

\section{Introduction}
\label{sec:intro}
Controlling the rotational motion of molecules with external electric fields is among the most
interesting goals in physical chemistry. The simplest approach to theoretically describe this
field-induced control relies on applying the rigid-rotor approximation. For many molecular species,
this approach has been shown to be sufficient~\cite{Hamilton:PRA72:043402, Rouzee:PRA73:033418,
   Rouzee:PRA77:043412, Trippel:PRA89:051401R, Trippel:PRL114:103003}, even for some floppy
molecules~\cite{Thesing:PRA98:053412}. However, the coupling of the overall angular momentum to
additional angular momenta or internal rotations cannot be neglected under certain circumstances.
For instance, it has been shown that coupling of nuclear spins and the overall angular momentum can
have a significant impact on the rotational dynamics on experimentally relevant
timescales~\cite{Altkorn:MP55:1, Yan:JCP98:6869, Cool:JCP103:3357, Zhang:JCP104:7027,
   Wouters:CP218:309, Sofikitis:JCP127:144307, Bartlett:PCCP11:142, Bartlett:PCCP12:15689,
   Grygoryeva:JCP147:013901}.

Fixing molecules in space, \ie, aligning and orienting them~\cite{Loesch:JCP93:4779,
   Friedrich:JCP111:6157, Stapelfeldt:RMP75:543}, is of particular interest among rotational control
schemes as it reduces the blurring of experimental observables caused by averaging over the random
orientations and allows to obtain information in the molecular frame~\cite{Spence:PRL92:198102,
   Filsinger:PCCP13:2076, Reid:PTRSA376:20170158}. Molecules can be aligned by subjecting them to
nonresonant laser fields. If the laser pulse is switched on slowly compared to the molecular
rotational period, adiabatic alignment is achieved. On the other hand, by using short laser pulses,
coherent superpositions of field-free rotational states are created. The wave packets rephase
periodically and show revivals of the alignment in field-free space~\cite{Seideman:PRL83:4971,
   RoscaPruna:JCP116:6579, Hamilton:PRA72:043402, Karamatskos:NatComm10:3364}. Such coherent
superpositions can also be obtained by shaped laser pulses which are turned off rapidly compared to
the rotational period of the molecule~\cite{Seideman:JCP115:5965, Underwood:PRL90:223001,
   Underwood:PRL94:143002, Goban:PRL101:013001, Chatterley:JCP148:221105}.

Recently, it has been shown that the impulsive alignment of \Itwo molecules can only be described
accurately if the nuclear-quadrupole coupling is taken into account~\cite{Thomas:PRL120:163202}. We
previously developed a generalized methodology to describe the rovibrational dynamics of molecules
including nuclear-quadrupole interactions~\cite{Yachmenev:JCP151:244118}. Here, we systematically
analyze the impact of this coupling in different existing techniques for laser alignment of
prototypical linear and asymmetric top molecules using \Itwo and 1,4-diiodobenzene as examples. Our
results thereby serve as a guideline to understand under which circumstances the quadrupole coupling
has a significant influence on the rational dynamics.

\section{Theoretical Description}
\label{sec:theory}
In \Itwo and 1,4-diiodobenzene (DIB), the interaction between the nuclear quadruple moments of two
equivalent iodine $^{127}$I nuclei and the electric field gradients, arising from the charge
distributions of the surrounding nuclei and electrons, lead to the well known hyperfine splittings
of the rotational energy levels. Each rotational level of the molecule is thus split into a maximum
of 36 sub-levels, labeled by the quantum number $F$ of the total angular momentum operator
$\boldsymbol{F}=\boldsymbol{J}+\boldsymbol{I}$, where $\boldsymbol{J}$ is the rotational angular
momentum neglecting the spin, which for the molecules considered here is the angular momentum of
overall rotation. Here, $\boldsymbol{I}=\boldsymbol{I}_{1}+\boldsymbol{I}_{2}$ is the collective
nuclear spin angular momentum operator with $I_1=I_2=5/2$ and thus $0\leq I\leq5$.

The theoretical model for the nuclear-quadrupole interactions has been described
before~\cite{Yachmenev:JCP151:244118}. Briefly, within the Born-Oppenheimer and semirigid-rotor
approximations, the field-free Hamiltonian can be written as
\begin{equation}
   \label{eqn:H_field_free}
   H_\text{mol} = H_\text{rot} + \sum_{l=1,2} \boldsymbol{V}(l) \cdot \boldsymbol{Q}(l),
\end{equation}
where $H_\text{rot}$ is the semirigid-rotor Hamiltonian. The rotational constant of \Itwo was
experimentally determined to $B=1118.63~\text{MHz}$~\cite{Thomas:PRL120:163202}. For DIB, the
rotational constants $B_z=A=5712.768~\text{MHz}$, $B_y=B=159.017~\text{MHz}$, and
$B_x=C=154.710~\text{MHz}$ were obtained from a geometry optimization using density functional
theory (DFT) with the B3LYP functional and the def2-QZVPP basis
set~\cite{Weigend:JCP119:12753,Weigend:PCCP7:3297}; for the iodine atoms the effective core
potential def2-ECP was used~\cite{Peterson:JCP119:11113}. All electronic structure calculations
employed the quantum-chemistry package ORCA~\cite{Neese:wircms2:73, Neese:wircms8:e1327}. The
molecule-fixed frame (MFF), $x,y,z$, is defined by the principal axes of inertia. In the second
term, $\boldsymbol{Q}(l)$ is the nuclear-quadrupole tensor of the $l$-th iodine nucleus and
$\boldsymbol{V}(l)$ is the electric-field-gradient tensor at the instantaneous position of the
corresponding nucleus. Due to the symmetry of DIB, the electric-field-gradient (EFG) tensors on the
two iodine centers are equal to each other with nonzero elements only on the diagonal
$V_{xx}=-5.5879~\au$, $V_{yy}=-6.2295~\au$ and $V_{zz}=11.8174~\au$. The nuclear quadrupole moment
for $^{127}$I is $Q=-696~\text{mb}$~\cite{Pyykko:MolPhys106:1965}. For \Itwo, instead of computing
the EFG tensors we used the experimental nuclear-quadrupole coupling constant
$\chi_{zz} = eQV_{zz}=-2.45258~\text{GHz}$~\cite{Yokozeki:JCP72:3796}, with the elementary charge
$e$.

The interaction of a molecule with a nonresonant laser field, linearly polarized along the
laboratory-fixed $Z$ axis, is given by
\begin{equation}
   H_\text{las} (t) = -\frac{I(t)}{2\varepsilon_0 c}\alpha_{ZZ}
   \label{eqn:H_las}
\end{equation}
where $\alpha_{ZZ}$ is the element of the polarizability tensor of the molecule along the laser
polarization axis. The polarizabilities defined in the laboratory frame are directly transformed to
the molecular frame $\alpha_{ij}$ ($i,j=x,y,z$)~\cite{Owens:JCP148:124102}. For both, \Itwo and DIB,
the polarizability tensor $\alpha_{ij}$ is diagonal in the inertial frame. For \Itwo, we used the
values $\alpha_{xx}=\alpha_{yy}=7.94~\text{\AA}^3$ and
$\alpha_{zz}=13.96~\text{\AA}^3$~\cite{Maroulis:JPCA101:953}. For DIB, calculated values of
$\alpha_{xx}=11.307~\text{\AA}^3$, \mbox{$\alpha_{yy}=16.676~\text{\AA}^3$}, and
$\alpha_{zz}=32.667~\text{\AA}^3$ were used~\footnote{For convenience and tradition, we specify
   polarizabilities in $\text{\AA}^3$, which can easily be converted to SI units as
   $1\text{\AA}^3=10^{6}~\text{pm}^3$.}.
Calculations of the EFG and polarizability tensors
for the DIB molecule were carried out at the DFT/B3LYP level of theory using the all-electron scalar
relativistic Douglas-Kroll-Hess Hamiltonian~\cite{Neese:JCP122:204107} with the DKH-def2-TZVP basis
set~\cite{Jorge:JCP130:064108,Campos:MolPhys111:167}.

To study the rotational dynamics of \Itwo and DIB, we solved the time-dependent Schr{\"o}dinger equation
(TDSE) for the full Hamiltonian
\begin{equation}
   H (t) = H_\text{mol} + H_\text{las}(t).
   \label{eqn:H_full}
\end{equation}
The time-dependent wave function was built from a superposition of the field-free spin-rotational
eigenfunctions of $H_\text{mol}$. The time-dependent coefficients were determined by
numerical solution of the TDSE using the split-operator method using
RichMol~\cite{Owens:JCP148:124102}.
To obtain the field-free eigenfunctions, we solved the time-independent
Schr{\"o}dinger equation for the Hamiltonian $H_\text{mol}$. The matrix representation of
$H_\text{mol}$ was constructed in a symmetry-adapted coupled basis $\ket{F,J,I,w}$ of the rotational
wave functions $\ket{J,w}$ and the nuclear-spin functions $\ket{I}$. Here, $w$ represents additional
rotational (pseudo) quantum numbers, such as $K_a$ and $K_c$ for DIB. For \Itwo, we took into
account the symmetry requirement that $J$ and $I$ have to be either both even or both
odd~\cite{Kroll:PRL23:631}, while for DIB all combinations of basis states are allowed. The
rotational states $\ket{J,w}$ were obtained as linear combinations of symmetric top functions by
diagonalizing $H_\text{rot}$. The explicit expressions for the matrix elements of the quadrupole
coupling Hamiltonian and various multipole Cartesian tensor operators can be found
elsewhere~\cite{Cook:AJP39:1433,Yachmenev:JCP147:141101,Yachmenev:JCP151:244118}. To obtain the
matrix representation of the interaction Hamiltonian, the matrix elements were first set up in the
coupled basis~\cite{Cook:AJP39:1433} and then transformed to the field-free
eigenbasis~\cite{Yachmenev:JCP151:244118}.
The alignment is quantified by \costhreeD, with the Euler angle
$\theta$ between the molecule-fixed $z$ and the laboratory-fixed $Z$ axes.

\section{Results and Discussion}
\label{sec:results}
\begin{figure*}
   \includegraphics[width=\linewidth]{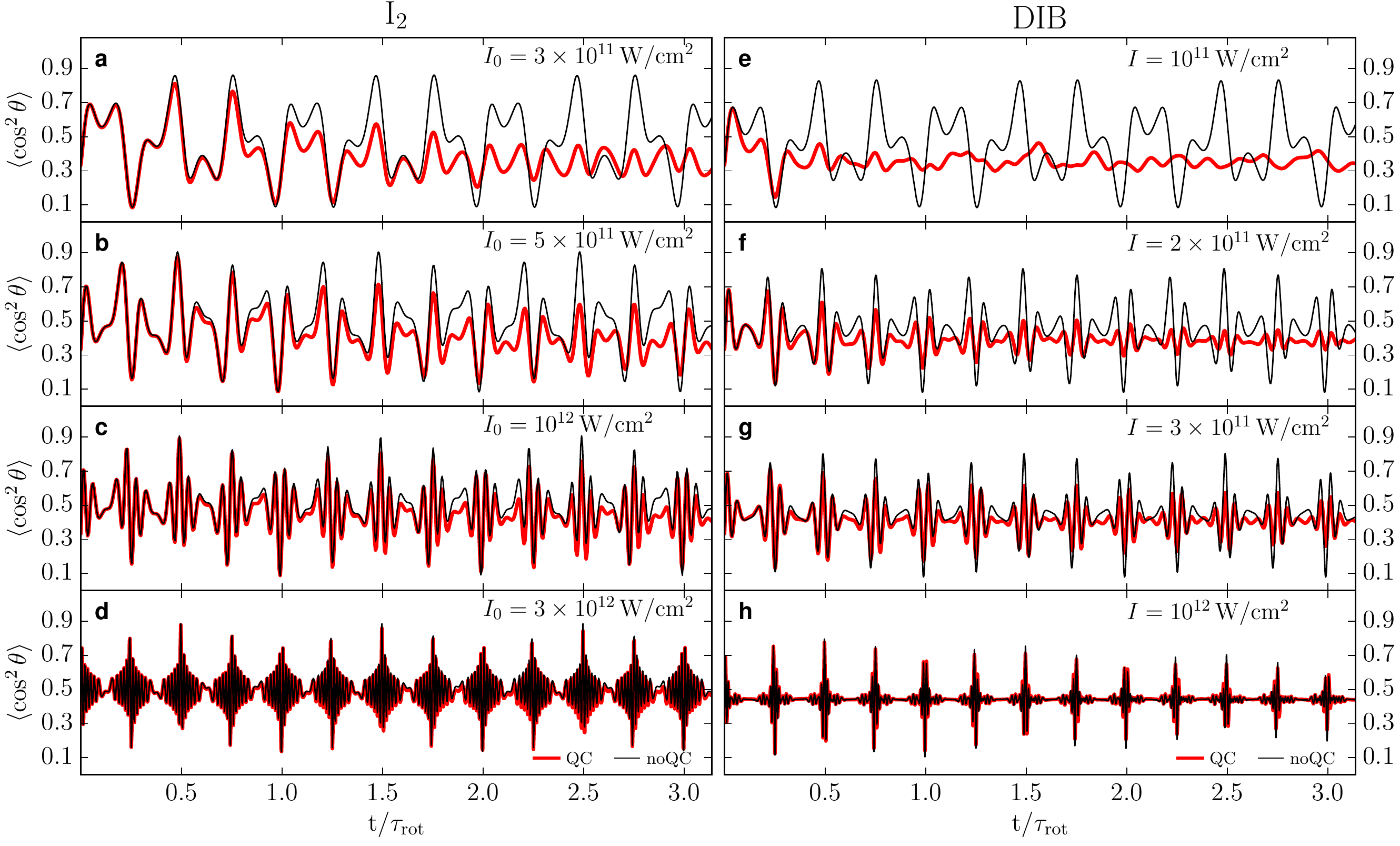}
   \caption{Impulsive alignment induced by laser pulses with $\tau_\text{FWHM}=1~\text{ps}$ for
     (a--d) \Itwo and (e--h) DIB including (QC) and neglecting the nuclear-quadrupole coupling
     (noQC). The expectation value \costhreeD is shown as a function of $t/\taurot$, where \taurot
     is the rotation period of the molecules. For each row, the laser intensities were chosen to
     create rotational wave packets involving similar distributions of $J$ values for both molecules
     when the quadrupole coupling is neglected.}
   \label{fig:results_impulsive}
\end{figure*}

\paragraph{Adiabatic alignment}
We considered a linearly polarized laser pulse with a Gaussian envelope
$I(t)=I_0 \exp\left(-4\ln2t^2/\tau_\text{FWHM}^2\right)$, where $\tau_\text{FWHM}=8~\text{ns}$ and
$I_0 = 2\times 10^{11}~\Wpcmcm$. For the rotational temperature $\Trot=0~\text{K}$, we assumed that
all hyperfine states corresponding to the rotational ground state $J=0$ were populated according to
their statistical weights. For \Itwo, the spin-rotational states with even $I$ have equal weights
while the ones with odd $I$ have zero weights~\cite{Kroll:PRL23:631}. For DIB, the Hydrogen
nuclear-spin functions result in weights of 7 (3) for even (odd) $I$; see \appautoref{sec:weights}.
The adiabatic alignment with nuclear-quadrupole coupling was computed by averaging over individually
obtained alignment results for the $15$ ($36$) initial nuclear-spin states of \Itwo (DIB).

The interaction of the external electric field with the polarizability of the molecules is much
stronger than the quadrupole-coupling interaction, resulting in the decoupling of the nuclear spins
from the overall-rotation angular momentum. As a result, the influence of the nuclear-quadrupole
coupling on the adiabatic alignment is negligible for both molecules. At the peak intensity, we thus
obtained $\costhreeD=0.946$ and $\costhreeD=0.988$ for \Itwo and DIB, respectively, both, including
and neglecting the quadrupole coupling.

\paragraph{Impulsive alignment}
We analyzed the impulsive alignment induced by short ($\tau_\text{FWHM}=1~\text{ps}$) nonresonant
linearly polarized laser pulses with Gaussian envelopes. We compared the post-pulse dynamics
including and neglecting the nuclear-quadrupole coupling for the rotational gound states, \ie,
$\Trot=0~\text{K}$. To allow for a better comparison of the alignment with and without the coupling,
we used initial states with well defined rotational quantum numbers, \ie, uncoupled states
$\ket{IM_I}\ket{00}$ ($\ket{IM_I}\ket{0_{00}0}$) for \Itwo (DIB), and averaged over the results for
the different spin isomers, see above. We point out that the corresponding field-free eigenstates
have contributions of $J>0$ rotational states, which might lead to additional differences in the
dynamics. To solve the TDSE, we projected each initial state onto the field-free eigenbasis.

\autoref{fig:results_impulsive} shows the post-pulse alignment for (a--d) \Itwo for intensities
$3\times10^{11}~\Wpcmcm<I_0<3\times10^{12}~\Wpcmcm$ and (e--h) DIB for intensities
$10^{11}~\Wpcmcm<I_0<10^{12}~\Wpcmcm$ as a function of time in units of the rotational periods,
$\taurot=446.98~\text{ps}$ and $\taurot=3187.48~\text{ps}$ for \Itwo and DIB, respectively. The
post-pulse alignment simulated without the quadrupole interaction shows typical revival structures
for rotational wave packets induced by short laser pulses. For the smallest intensities,
\autoref[a,~e]{fig:results_impulsive}, the field-dressed dynamics is dominated by a few low-energy
rotational states and \costhreeD oscillates with the period \taurot. For DIB and
$I_0=10^{11}~\Wpcmcm$, contributions of rotational states with $K_a>0$ are negligible and the
post-pulse dynamics is similar to the one of the linear molecule \Itwo. With increasing intensity
$I_0$, more highly excited rotational states are involved in the dynamics. The rotational dynamics
of DIB in \autoref[h]{fig:results_impulsive} shows a decrease of the peak alignment over time
resulting from the asymmetry splitting of rotational states with $K_a>0$~\cite{Rouzee:PRA73:033418,
   Holmegaard:PRA75:051403R}.

The effect of the nuclear-quadrupole coupling on the alignment depends strongly on the laser
intensity. For the low intensities in \autoref[a,~e]{fig:results_impulsive} it is strongest and the
field-free alignment decreases over time for both \Itwo and DIB. However, during the short pulse the
quadrupole interaction plays a negligible role for both molecules due to a decoupling of the
nuclear spin and overall-rotation angular momenta. The laser field only affects
the rotational part of the wave packet, leaving the nuclear-spin quanta unchanged. As a
consequence, the alignment traces with and without the
quadrupole coupling are very similar to each other directly after the laser pulse. This also holds
for higher laser intensities.

For DIB, the alignment then decreases quickly after the pulse, while for \Itwo the rotational
dynamics starts to differ at $t\approx\taurot/2$. This decrease in the field-free alignment for
small intensities can be rationalized in terms of the hyperfine energy levels of the
Hamiltonian~\eqref{eqn:H_field_free}. For the low-energy rotational states, which contribute most to
the dynamics in \autoref[a,~e]{fig:results_impulsive}, the hyperfine-splitting patterns depend
strongly on $J$. These irregular patterns introduce incommensurate frequencies that lead to a
dephasing of the wave packet, thus preventing strong revivals of the alignment. In contrast to
\Itwo, for DIB the energy differences of hyperfine components within the low-energy rotational
states are similar to the energy differences between the rotational levels themselves and strong
inter-$J$ coupling is observed in the spin-rotational states. As a consequence, the alignment of DIB
is affected much more strongly and faster, with respect to the rotational timescale, than it is for
\Itwo.

The influence of the quadrupole coupling can also be interpreted using a classical picture. The
precession of $\boldsymbol{I}$ and $\boldsymbol{J}$ around $\boldsymbol{F}$ results in a variation
of their projections $M_J$ and $M_I$, leading to a decrease of the alignment. In addition, the
magnitudes $|\boldsymbol{I}|$ and $|\boldsymbol{J}|$ are changed over time by the quadrupole
coupling, further affecting the rotational dynamics. Note that for \Itwo, the changes in
$|\boldsymbol{J}|$ are very small but a change in $|\boldsymbol{I}|$ stills affects the spatial
orientation of $\boldsymbol{J}$, since the total angular momentum $\boldsymbol{F}$ is preserved.

As the laser intensity was risen, \autoref[b--d,~f--h]{fig:results_impulsive}, the influence of the
quadrupole interaction diminished for both molecules. In \autoref[d,~h]{fig:results_impulsive},
minor differences can only be observed after the first rotational period. For these strong fields,
highly excited states with up to $J\approx44$ (and $K_a\approx 10$ for DIB) dominate the post-pulse
dynamics, for which the hyperfine patterns become increasingly uniform~\cite{Gordy:MWMolSpec,
   Kroto:MolecularRotationSpectra}. 
For these large $J$, the matrix elements of $\cos^2\theta$ that
contribute significantly to the alignment are those between field-free eigenstates with
$\Delta{F}=\Delta{J}$ and the same nuclear-spin contributions. Since the hyperfine energy shifts are
approximately the same for these states, their energy gaps are very similar to those between the
corresponding rotational levels. As a result, we observe only a very weak dephasing in
\autoref[d,~h]{fig:results_impulsive}.

If field-free eigenstates are used as initial states, the averaged alignment results for \Itwo do not
differ significantly from the result in \autoref{fig:results_impulsive}. For DIB on the other hand,
we observe small deviations originating from a considerable mixing of different $J$-states in the
hyperfine eigenstates, modifying the initial rotational wave function. The impact of the quadrupole
interaction is, however, qualitatively the same. The different sets of initial states can only be
considered equivalent when the result is averaged over all different spin isomers and $M$-states for
a given rotational level. Individual eigenstates are in general linear combinations of uncoupled
states with different values of $I$, $M_I$ and $M_J$, even if the coupling of different $J$-states
by the quadrupole interactions is neglected.

\paragraph{Truncated pulse alignment}
\begin{figure}
   \includegraphics[width=\linewidth]{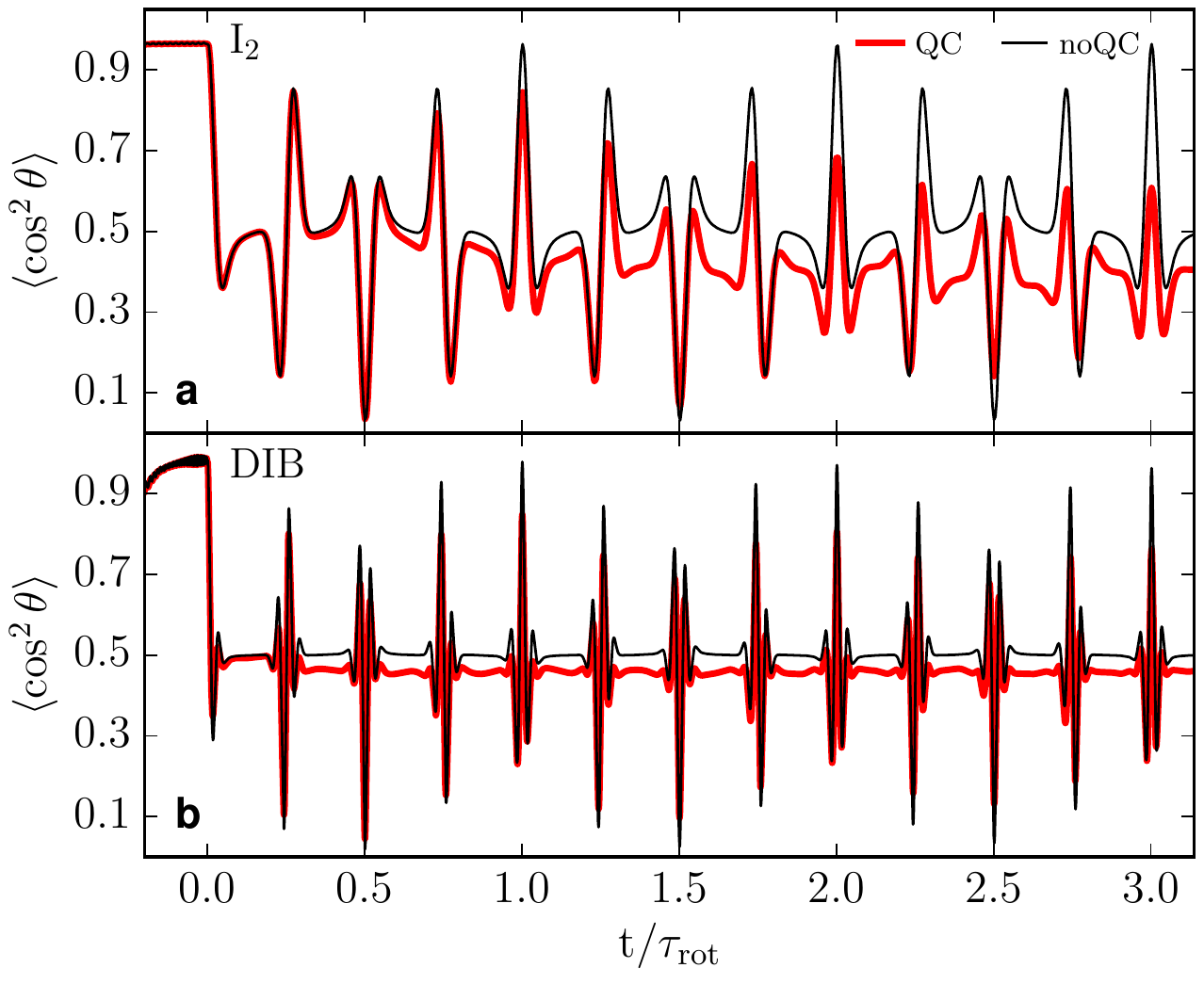}
   \caption{Truncated pulse alignment with and without quadrupole coupling (QC and noQC) as a
      function of $t/\taurot$ for (a) \Itwo and (b) DIB following a truncated laser pulse with a
      maximum laser intensity of $I_0=5\times10^{11}\Wpcmcm$. The rising and falling edges of the
      pulse had Gaussian shapes with $\tau_\text{FWHM}=600~\text{ps}$ and $2~\text{ps}$,
      respectively.}
   \label{fig:results_truncated}
\end{figure}
For asymmetric top molecules, truncated laser pulses typically allow one to obtain larger degrees of
field-free alignment than short laser pulses, taking advantage of the initial adiabatic
alignment~\cite{Seideman:JCP115:5965, Chatterley:JCP148:221105}. Here, we consider a laser pulse
with a rising and falling edge, both with a Gaussian shape corresponding to
$\tau_\text{FWHM}=600~\text{ps}$ and $\tau_\text{FWHM}=2~\text{ps}$, respectively, and a peak
intensity of $I_0=5\times10^{11}\Wpcmcm$.

The alignment results for \Itwo and DIB as a function of $t/\taurot$ are shown in
\autoref[a,~b]{fig:results_truncated}, respectively. For both molecules, strong alignment was
reached before the cut-off of the laser field with \costhreeD close to $1$. As in the adiabatic
regime, the effect of the quadrupole coupling on the degree of alignment is very weak in the
presence of the field. After the laser field is switched off, the quadrupole-free alignment shows a
typical revival structure. For the linear \Itwo molecules, \costhreeD at the full revival
$t=\taurot$ reaches the same value as during the pulse, $\costhreeD=0.96$. Analogous behavior was
observed for DIB with $\costhreeD=0.98$. Note that such high degrees of the post-pulse alignment for
DIB are generally possible because the molecule is a near symmetric-top with only small populations
of states with $K_a>0$, effectively reducing the dynamcis to that of a linear rotor. Generally, the
peak alignment of asymmetric top molecules at the full revival does not reach the same value as at
the peak intensity of the truncated pulse~\cite{Underwood:PRL94:143002, Chatterley:JCP148:221105}.

For both molecules, the dephasing due to the nuclear-quadrupole coupling is comparable to the
impulsive alignment case with intermediate intensities, see \autoref[b,~f]{fig:results_impulsive}.
For DIB, the inter-$J$ coupling due to quadrupole interaction noticeably influences the populations
of rotational states during the pulse, where the dynamics shows nonadiabatic behavior. The alignment
then starts to slightly deviate from the quadrupole-free results directly after the truncation of
the laser field. By including the coupling, the strongest peak alignment was observed at the full
revival $t=\taurot$ with $\costhreeD=0.85$ ($0.84$) for \Itwo (DIB), which is larger than the peak
alignment obtained in the impulsive regime, see \autoref[d,~h]{fig:results_impulsive}. However, the
enhancement of the post-pulse alignment by using truncated laser pulses instead of impulsive-kick
pulses was much smaller than without the quadrupole interaction.

\paragraph{Post-pulse dynamics of excited rotational states}
\begin{figure}
   \centering%
   \includegraphics[width=\linewidth]{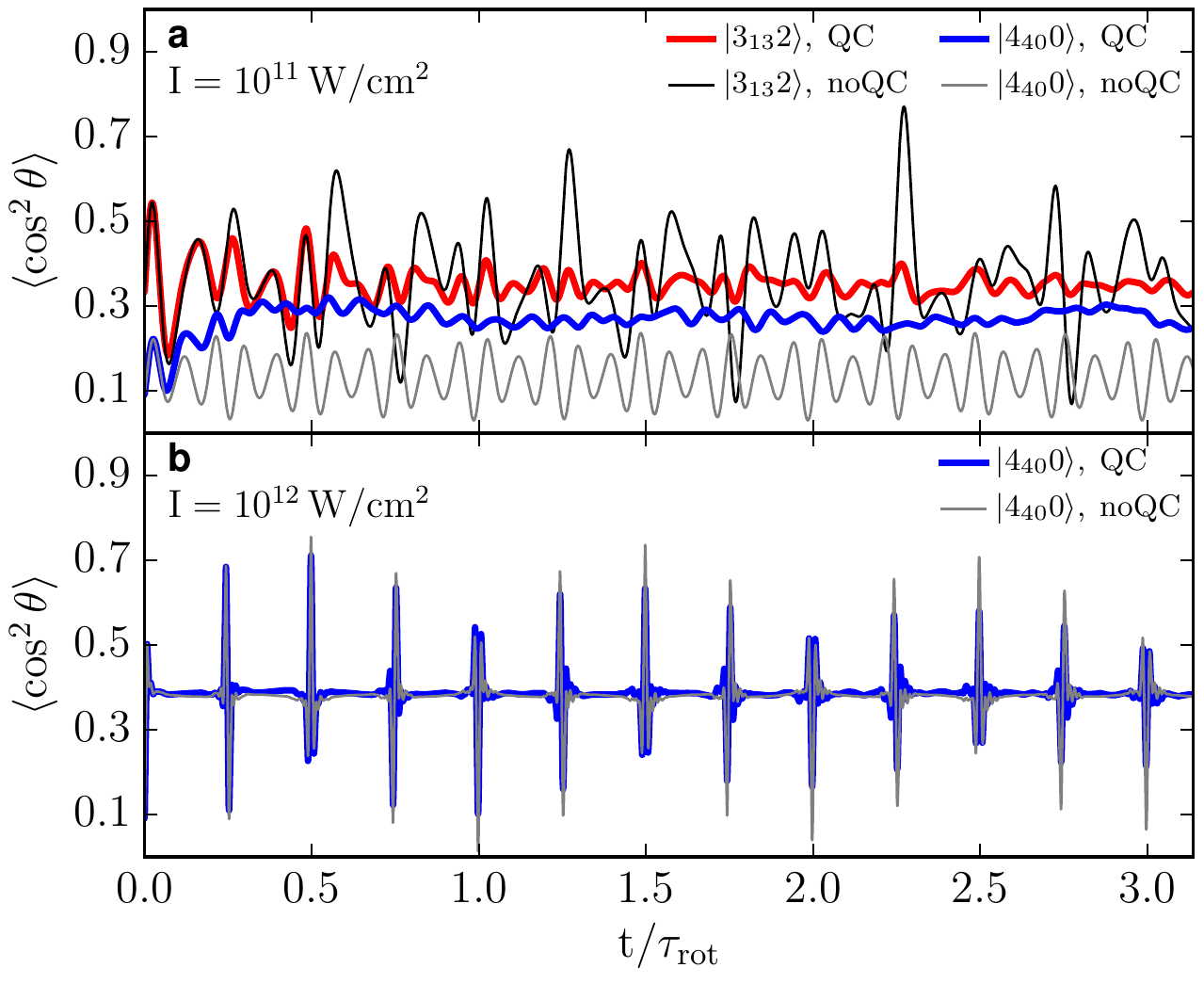}
   \caption{Impulsive alignment for the initial states $\ket{3_{13}2}$ and $\ket{4_{40}0}$ of DIB
      including the quadrupole coupling (QC) and neglecting it (noQC) as a function of $t/\taurot$.
      The duration of the laser pulse is $\tau_\text{FWHM}= 1~\text{ps}$ and (a)
      $I_0=1\times10^{11}\Wpcmcm$ and (b) $I_0=1\times10^{12}\Wpcmcm$.}
   \label{fig:results_excited}
\end{figure}
While state-selected molecular beams~\cite{Filsinger:PCCP13:2076, Hutzler:CR112:4803,
   Meerakker:CR112:4828, Chang:IRPC34:557, Nielsen:PCCP13:18971, Horke:ACIE53:11965,
   Karamatskos:NatComm10:3364} or ultracold-molecules techniques~\cite{Ulmanis:CR112:4890,
   Quemener:CR112:4949} can produce near-0~K samples, this is not generally feasible, especially not
for the heavy organic molecules discussed here. Therefore, we investigated the impulsive-alignment
dynamics for several initially excited states of \Itwo and DIB. Since comparable results were
obtained for both molecules, we focus our analysis on DIB, namely the rotational states
$\ket{J_{K_aK_c},M_J}=\ket{3_{13},2}$ and $\ket{4_{40},0}$. For both excited states, we observe a
qualitatively similar influence of the quadrupole coupling as for the rotational ground state. For
$I_0=1\times10^{11}\Wpcmcm$, \costhreeD in \autoref[a]{fig:results_excited} approaches $1/3$ on
timescales similar to the one in \autoref[e]{fig:results_impulsive}. For the higher intensity
$I_0=1\times10^{12}\Wpcmcm$, we find a more significant decrease of the post-pulse alignment for the
initial state with $J=4,M_J=0$, \autoref[b]{fig:results_excited} than for the rotational ground
state, \autoref[h]{fig:results_impulsive}. This can be attributed to the populations of field-free
eigenstates, which are shifted toward smaller values of $J$ for \ket{4_{40},0}. However, for other
excited states the impact of the quadrupole coupling for higher laser intensities can be smaller, as
is the case for $\ket{3_{13}2}$. Thus, for a thermal ensemble of molecules, we expect a small
influence of nuclear-quadrupole interactions on the post-pulse alignment induced by strong laser
fields. However, to accurately describe the alignment the coupling cannot be fully neglected even in
this regime. Furthermore, on longer timescales even small frequency shifts will lead to a
significant decrease of the alignment~\cite{Thomas:PRL120:163202}.

\section{Summary and Conclusions}
\label{sec:summary}
In conclusion, a significant dephasing of rotational wave packets was observed in the post-pulse
dynamics for different laser-field shapes and intensities. The influence on the degree of alignment
is the largest if low-energy, small-$J$ rotational states dominate the dynamics and diminishes for
highly excited states. For initially excited rotational states, the quadrupole coupling has a
similar effect on the post-pulse dynamics as for the rotational ground state, which we expect to
hold for thermal ensembles as well. Adiabatic alignment is essentially not affected by the
nuclear-quadrupole interactions.

Our results emphasize the need to take into account the nuclear-quadrupole interactions in the
description of field-free alignment for molecules with heavy nuclei with large nuclear quadrupoles.
Since many biomolecules include such heavy elements, we plan to investigate other molecular species
and their properties including rotational constants and molecular symmetry in the context of
nuclear-quadrupole interactions.

\appendix
\part*{\normalsize Appendix}
\section{Spin-statistical weights of hyperfine states}
\label{sec:weights}
To derive the weights of the iodine-spin-rotational wave functions of DIB, we made use of its
molecular symmetry group $D_\text{2h}$, the corresponding character table can be found
elsewhere~\cite{Bunker:MolecularSymmetry}. Pure rotational states obey either $A_\text{g}$,
$B_\text{1g}$, $B_\text{2g}$ or $B_\text{3g}$ symmetry, while the iodine nuclear-spin functions
$\ket{I}$ obey $A_\text{g}$ or $B_\text{1u}$ symmetry for $I$ odd or even, respectively. This
follows from the effect of symmetry operations $\hat{P}$ that involve a permutation of the iodine
nuclei, $\hat{P}\ket{I}=(-1)^{I_1+I_2+I}\ket{I}$, yielding a sign change for even values of $I$. The
representation generated by the hydrogen nuclear-spin functions was
derived~\cite{Bunker:MolecularSymmetry} as
$7A_\text{g}\oplus 3B_\text{1g}\oplus 3B_\text{2g}\oplus 3B_\text{3g}$. We considered rotational
states having either $A_\text{g}$ ($\ket{J_{K_aK_c}}=\ket{0_{00}}$ and $\ket{4_{40}}$) or
$B_\text{2g}$ ($\ket{3_{13}}$) symmetry. With the requirement that the total internal wave function
has to be of $B_\text{1g}$ or $B_\text{1u}$ symmetry, we obtained weights of 7 (3) for even (odd)
$I$ for the former case, while for the latter all spin-rotational states have equal weights.

In the case of the diatomic \Itwo molecule, there are no additional nuclear spins leading to
degeneracies of the hyperfine levels. However, in the electronic and vibrational ground state
considered here, states with $J$ and $I$ having opposite parities are forbidden due to the
generalized symmetrization postulate~\cite{Kroll:PRL23:631}.

\begin{acknowledgments}
   We thank Andrey Duchko for his contributions in an early part of the project.

   This work has been supported by the Deutsche Forschungsgemeinschaft (DFG) through the priority
   program ``Quantum Dynamics in Tailored Intense Fields'' (QUTIF, SPP1840, KU~1527/3, YA~610/1) and
   through the clusters of excellence ``Center for Ultrafast Imaging'' (CUI, EXC~1074, ID~194651731)
   and ``Advanced Imaging of Matter'' (AIM, EXC~2056, ID~390715994). R.G.F.\ gratefully acknowledges
   financial support by the Spanish Project No. FIS2017-89349-P (MINECO), the Andalusian research
   group FQM-207, the Consejer\'{\i}a de Conocimiento, Investigaci\'on y Universidad, Junta de
   Andaluc\'{\i}a, and the European Regional Development Fund (ERDF, SOMM17/6105/UGR).
\end{acknowledgments}

\bigskip
\def\bibsection{}
\textbf{References}
\medskip
\bibliographystyle{achemso}
\bibliography{string,cmi}

\end{document}